\documentclass[aps,prl,reprint,superscriptaddress]{revtex4-2}

\usepackage[utf8]{inputenc}
\usepackage[T1]{fontenc}
\usepackage[english]{babel}
\usepackage{mathtools}
\usepackage{bm}
\usepackage{amsfonts}
\usepackage{nicefrac}

\usepackage[colorlinks=true,
    citecolor=blue,
    linkcolor=blue,
    urlcolor=blue]{hyperref}
\usepackage{cleveref}
\usepackage{orcidlink}
\usepackage{mathptmx}
\usepackage{xcolor}

\begin{document}

\title{Evidence for renormalized instantons in real-time simulations of vacuum decay}

\author{Emilie Hertig\,\orcidlink{0000-0001-9189-4035}}
\altaffiliation{Corresponding author}
\email{emh83@cam.ac.uk}
\affiliation{Institute of Astronomy, University of Cambridge, Madingley Road, Cambridge CB3 0HA, UK}
\affiliation{Kavli Institute for Cosmology, University of Cambridge, Madingley Road, Cambridge CB3 0HA, UK}

\author{Alexander~C.~Jenkins\,\orcidlink{0000-0003-1785-5841}}
\email{acj46@cam.ac.uk}
\affiliation{Kavli Institute for Cosmology, University of Cambridge, Madingley Road, Cambridge CB3 0HA, UK}
\affiliation{DAMTP, University of Cambridge, Wilberforce Road, Cambridge CB3 0WA, UK}

\author{Hiranya~V.~Peiris\,\orcidlink{0000-0002-2519-584X}}
\affiliation{Kavli Institute for Cosmology, University of Cambridge, Madingley Road, Cambridge CB3 0HA, UK}
\affiliation{Institute of Astronomy, University of Cambridge, Madingley Road, Cambridge CB3 0HA, UK}
\affiliation{The Oskar Klein Centre for Cosmoparticle Physics, Department of Physics, Stockholm University, AlbaNova, Stockholm SE-106 91, Sweden}

\author{Andrew~Pontzen\,\orcidlink{0000-0001-9546-3849}}
\affiliation{Institute for Computational Cosmology, Department of Physics, Durham University, South Road, Durham, DH1 3LE, UK}

\author{Matthew C. Johnson\,\orcidlink{0000-0003-4199-0314}}
\affiliation{Perimeter Institute for Theoretical Physics, 31 Caroline St N, Waterloo, ON N2L 2Y5, Canada}
\affiliation{Department of Physics and Astronomy, York University, Toronto, ON M3J 1P3, Canada}

\begin{abstract}
    While vacuum decay is traditionally described by Euclidean instanton methods, lattice simulations enable real-time modeling of dynamical observables relevant to cosmology and upcoming cold-atom analog experiments. We investigate the relationship between these approaches by extracting ensemble-averaged bubble profiles from zero-temperature simulations of a relativistic scalar field. Our observed profiles differ markedly from the bare Coleman bounce and classical thermal predictions. However, we find that instanton solutions in an appropriately renormalized potential reproduce both the measured profiles and their dependence on the UV cutoff, and predict decay rates consistent with simulations across the parameter range considered. The fact that a single renormalized Euclidean object captures these independent observables provides strong evidence that renormalization accounts for the discrepancy between the two formalisms, and establishes a quantitative link between instanton predictions, lattice simulations, and forthcoming empirical tests of vacuum decay.
\end{abstract}

\date{\today}
\maketitle


\emph{Introduction---}%
False vacuum decay (FVD), a first-order phase transition in which bubbles of a new phase nucleate and expand in a metastable background, is a core feature of various quantum field theories (QFTs) and cosmological scenarios. With applications ranging from eternal inflation~\cite{guth_could_1983,aguirre_eternal_2007} to Higgs stability~\cite{buttazzo_investigating_2013,markkanen_cosmological_2018}, the study of FVD provides deep insight into the origin, evolution, and ultimate fate of the Universe. 

Recent advances in cold-atom analog experiments~\cite{fialko_fate_2015,braden_towards_2018,Billam:2020xna,jenkins_generalized_2023}, alongside forthcoming cosmic microwave background~\cite{ade_simons_2019,feeney_forecasting_2015} and gravitational-wave~\cite{kosowsky_gravitational_1992, kamionkowski_gravitational_1994,Caprini:2015zlo} observations, offer new prospects for probing aspects of FVD empirically. These developments highlight the need for a robust quantitative theoretical framework. The standard formalism developed by Coleman~\cite{coleman_fate_nodate,callan_fate_1977}, based on Euclidean instanton methods, establishes predictions for the decay rate and for the field configuration at nucleation (which we refer to as the bubble profile). However, this approach lacks a real-time description of the decay process and relies on symmetry assumptions which limit its applicability to more general settings, including multi-bubble configurations that are most relevant for observational tests~\cite{pirvu_bubble_2022}.

Real-time lattice simulations based on the truncated Wigner approximation have recently emerged as an alternative approach to modeling FVD dynamics~\cite{braden_new_2019}. This method consists of sampling initial fluctuations in the false vacuum and evolving the resulting field configuration classically. In the context of thermal vacuum decay, such simulations have revealed a rich phenomenology including clustering effects, oscillon precursors, and nonzero bubble velocities~\cite{pirvu_bubble_2022,pirvu_bubble_2023,pirvu_thermal_2024, pirvu_thermal_2024_b}. While nucleation events have also been successfully realized in the zero-temperature case, corresponding decay rates were found to exceed instanton predictions by several orders of magnitude~\cite{braden_erratum_2019}. This raises the question of whether these classically-allowed decay trajectories capture the same physical process as Coleman's tunneling picture~\cite{hertzberg_quantitative_2020,tranberg_bubble_2022}.

A possible explanation for the discrepancy lies in renormalization effects: short-wavelength fluctuations modify the effective potential experienced by long-wavelength modes on the lattice~\cite{braden_mass_2023}, which in turn governs the nucleation rate. Although corrections to the false and true vacuum masses can be computed analytically and are qualitatively consistent with higher decay rates, the lack of constraints on the full form of the effective potential has so far prevented a quantitative assessment of whether renormalization can account for the discrepancy.

In this \emph{Letter}, we address this question by exploring beyond decay rates and investigating the properties of individual bubbles in lattice simulations of a single scalar field in 1+1 dimensions. Building on analysis techniques previously applied to thermally-activated FVD~\cite{pirvu_bubble_2023,pirvu_thermal_2024_b}, we extract ensemble-averaged bubble profiles and find that they differ markedly from instanton predictions in the bare potential, exhibiting a clear dependence on the UV cutoff and providing evidence for renormalization. Furthermore, the profiles are accurately described by Coleman bounce solutions in an effective potential where masses are fixed analytically and a single free parameter is fitted. Decay rates predicted using these profile-constrained potentials are in quantitative agreement with those extracted from simulations, up to factors of order unity expected from prefactor uncertainties.

Our findings suggest that Euclidean and real-time descriptions of FVD are reconciled once renormalization is properly accounted for. This provides a quantitative foundation for connecting instanton predictions to both lattice simulations and forthcoming experimental investigations, while revealing new conceptual insights into the role of classically-allowed field trajectories in nonperturbative QFT phenomena.


\emph{Lattice simulations---}%
Our analysis focuses on the evolution of a scalar field in 1+1 dimensions with metastable potential
\begin{equation}\label{eq:Vbare}
    V_{\text{bare}}(\phi)=m_0^2\phi_0^2\left[1-\cos\left(\frac{\phi}{\phi_0}\right)+\frac{\lambda^2}{2}\sin^2\left(\frac{\phi}{\phi_0}\right)\right],
\end{equation}
which we model using the semiclassical stochastic lattice framework described in Ref.~\cite{braden_new_2019}. (We use units with $c=\hbar=1$ throughout this work.) True and false vacua are located at $\phi_{\text{TV}}=2n\pi\phi_0$ and $\phi_{\text{FV}}=\left(2n+1\right)\pi\phi_0$ with $n\in\mathbb{Z}$; $m_0$ and $\lambda$ are tunable parameters controlling the energy scale and barrier height of the potential, respectively, while $\phi_0$ sets the relative strength of fluctuations. Similar forms for $V_{\text{bare}}(\phi)$ were previously used in Refs.~\cite{braden_new_2019, pirvu_bubble_2023} and will be targeted by upcoming cold-atom analog experiments~\cite{fialko_fate_2015,braden_towards_2018}.

We approximate the false vacuum state by sampling stochastic fluctuations of a free massive scalar with effective mass $m_{\text{eff,FV}}$. Initial configurations for the field $\phi$ and its conjugate momentum $\Pi=\partial_t\phi$ are then given by
    \begin{align}
    \label{eq:phi_init}
    \phi(t=0,\,x)&=\phi_{\text{FV}}+\frac{1}{\sqrt{L}}\sum_{n\neq0}^{|n|\leq n_{\text{cut}}}{\frac{\hat{\alpha}_n}{\sqrt{2\omega_n}}}e^{ik_nx},\\
    \label{eq:pi_init}
    \mathrm{\Pi}(t=0,\,x)&=\frac{1}{\sqrt{L}}\sum_{n\neq 0}^{|n|\leq n_{\text{cut}}}{\sqrt{\frac{\omega_n}{2}}\hat{\beta}_ne^{ik_nx}},
    \end{align}
where $L$ represents the size of our periodic box, and $\omega_n^2=m_{\textrm{eff,FV}}^2+k_n^2$ for a wavevector $k_n=2\pi n/L$. The factors $\hat{\alpha}_n$ and $\hat{\beta}_n$ are independent, unit-variance complex Gaussian random variables such that $\hat{\alpha}_n=\hat{\alpha}_{-n}^{*}$ and $\hat{\beta}_n=\hat{\beta}_{-n}^{*}$. For a given number of lattice points $n_{\text{lat}}$, we populate Fourier modes of the field up to a UV cutoff $n_{\textrm{cut}} \leq n_{\text{lat}}/2$. Time evolution is then computed by solving the classical Hamiltonian equations of motion; this procedure is implemented in the pseudospectral code presented in Refs.~\cite{braden_new_2019,jenkins_analog_2024}.

In this work, we consider a box size $m_0L=200$ divided into $n_{\text{lat}}=4096$ lattice points and use the same spacing in the time direction. Simulations are terminated at $m_0t=4000$ or once 50\% of the spatial volume has transitioned to the true vacuum, whichever occurs first. We probe renormalization effects by varying the UV cutoff $n_{\text{cut}}$ between 128 and 1024, and keep the bare potential parameters fixed at $m_0=1$, $\lambda=1.6$ and $\phi_0=1.8$. These values were empirically selected to give inverse decay rates comparable to simulation timescales in the chosen $n_{\text{cut}}$ range.


\emph{Renormalized potential---}%
The short-wavelength fluctuations explicitly realized in our simulations modify the dynamics of IR modes, which are the relevant degrees of freedom in the semiclassical instanton description. Lattice results should therefore be compared to Euclidean predictions using a renormalized effective potential $V_{\text{eff}}$~\footnote{Note that the effective potential here differs from the one-particle-irreducible (1PI) effective potential usually considered in path integral treatments of QFT, which is convex and thus does not admit bounce solutions~\cite{Peskin:1995ev}. While the 1PI effective potential describes the lowest-energy equilibrium state at fixed mean field $\bar{\phi}$, we are instead interested in the false vacuum, which is an excited nonequilibrium state of the theory. We operationally define $V_\mathrm{eff}$ here as the input to Euclidean bounce calculations that successfully accounts for renormalization effects.}.

Following Ref.~\cite{braden_mass_2023}, we compute the renormalized true- and false-vacuum masses by resumming Gaussian fluctuations about the instantaneous mean field $\bar{\phi}$ (see End Matter). The fluctuation spectra in Eqs.~\eqref{eq:phi_init} and~\eqref{eq:pi_init} are initialized using the effective false-vacuum mass $m_{\mathrm{eff,FV}}$.

Although these calculations fix the local curvature of $V_{\mathrm{eff}}$ near the vacua, they do not uniquely determine its full shape. We therefore model $V_{\mathrm{eff}}$ using a three-term cosine expansion, treating the dimensionless energy splitting $\Delta\tilde{V}$ as the only free parameter and determining it by fitting the observed bubble profiles. Further details are given in the End Matter.


\emph{Bubble profiles---}%
The field profile at the time of bubble nucleation provides direct insight into the mechanism governing vacuum decay. We therefore compare simulation results with the Coleman bounce solution $\phi_b(x)$ determined by
\begin{equation}\label{eq:EOM}
\frac{\partial^2\phi_b}{\partial x^2}
+\frac{1}{x}\frac{\partial\phi_b}{\partial x}
=V'(\phi_b),
\end{equation}
where $V$ denotes either $V_{\text{bare}}$ or $V_{\text{eff}}$. Boundary conditions are given by $\partial_x \phi_b(0)=0$ and $\phi_b(x\to\pm\infty)=\phi_{\mathrm{FV}}$. The friction term in Eq.~\eqref{eq:EOM} arises from the $O(2)$ symmetry of the 1+1D Euclidean instanton~\cite{coleman_fate_nodate}. We also consider the critical classical bubble $\phi_c(x)$, which is a static solution of the real-time equations of motion and satisfies
\begin{equation}\label{eq:thermal_EOM}
\frac{\partial^2\phi_c}{\partial x^2}
=V'(\phi_c),
\end{equation}
with the same boundary conditions as before~\cite{linde_fate_1981,linde_decay_1983}. Although Eq.~\eqref{eq:thermal_EOM} is usually derived as a high-temperature limit of thermally-activated decay (which should be absent in our zero-temperature simulations), it more generally corresponds to the saddle-point configuration separating collapsing and expanding bubbles under classical evolution~\cite{linde_decay_1983}. Observing this profile would therefore suggest that the real-time decay channel realized in our simulations fundamentally differs from the process described by the Coleman instanton approach. To probe the impact of renormalization, we assess whether the measured profiles are better described by theoretical predictions evaluated in the bare lattice potential $V_{\text{bare}}$ or in the effective potential $V_{\text{eff}}$ for a best-fit value of $\Delta\tilde{V}$.

\begin{figure}[b]
    \centering
    \includegraphics[width=\columnwidth]{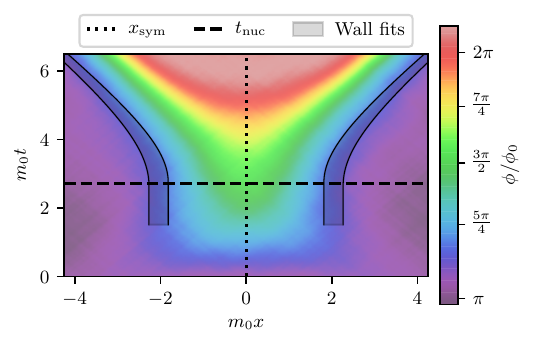}
    \caption{
        Ensemble-averaged bubble obtained from a stack of 500 deboosted realizations with $n_{\text{cut}}=512$.
        The shaded region indicates the range of outer wall trajectories used to determine the nucleation time $t_{\text{nuc}}$ (dashed line). The dotted line represents the symmetry axis of the late-time true-vacuum region.
    }
    \label{fig:average_bubble}
\end{figure}

\begin{figure*}[t]
\centering
\includegraphics[width=\textwidth]{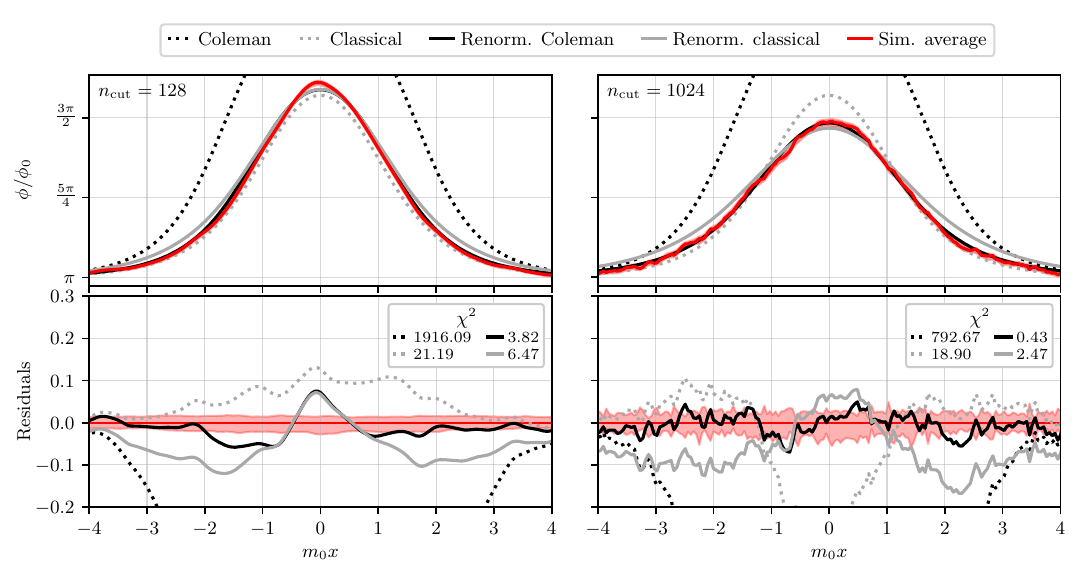}
\caption{Top: Ensemble-averaged bubble profiles (solid red curves) extracted from stacks of 500 deboosted realizations with $n_{\text{cut}}=128$ (left) and $n_{\text{cut}}=1024$ (right). Shaded bands indicate the 16th–84th percentile range obtained from 1000 bootstrap resamplings. Measured profiles are compared with Coleman bounce solutions [Eq.~\eqref{eq:EOM}, black] and critical classical bubbles [Eq.~\eqref{eq:thermal_EOM}, gray]. In each case, dotted curves correspond to predictions obtained from the bare potential, while solid curves denote renormalized solutions computed using the best-fit effective potential. Bottom: Residuals between the measured profiles and the corresponding theoretical predictions, with legends reporting reduced weighted $\chi^2$ values for each model.}
\label{fig:profiles}
\end{figure*}

Figure~\ref{fig:average_bubble} shows an ensemble-averaged field configuration obtained by transforming individual bubbles to their approximate rest frames and stacking. The bubble profile is then extracted at the onset of hyperbolic wall expansion (see End Matter and Ref.~\cite{PRD_companion}).

Our results for the smallest and largest UV cutoffs considered in this work ($n_{\text{cut}}=128$ and $n_{\text{cut}}=1024$) are shown in the left and right panels of Fig.~\ref{fig:profiles}, respectively. Bubble profiles obtained from simulations (red curves) exhibit a clear dependence on $n_{\text{cut}}$, with the central field amplitude decreasing as more short-wavelength fluctuations are included. Results for intermediate cutoff values, presented in our companion paper~\cite{PRD_companion}, confirm the same trend. This behavior constitutes direct evidence that the bubble nucleation process is significantly impacted by renormalization effects.

As a consequence, theoretical predictions obtained in the bare lattice potential fail to reproduce the observed field configurations. In particular, the bare Coleman bounce (dotted black curves) systematically overestimates the central field value, while the bare classical bubble (dotted gray curves) changes from slightly underestimating it in the left panel to significantly overestimating it at high $n_{\text{cut}}$.

We then compare simulation results with solutions of Eqs.~\eqref{eq:EOM} and~\eqref{eq:thermal_EOM} calculated in the effective potential $V_{\mathrm{eff}}$, varying the free parameter $\Delta\tilde{V}$ to obtain the best-fit profile in each case. The fit quality is quantified using a weighted $\chi^2$ statistic with weights proportional to $(\phi_{\text{sim}}(x)-\phi_{\text{FV}})^2/\sigma^2(x)$, where the variance $\sigma^2(x)$ is obtained from the bootstrap distribution symmetrized around the bubble center. This weighting scheme emphasizes the bubble interior and wall, which contain most of the information distinguishing the theoretical profiles; it also downweights the asymptotic false-vacuum tails, thus suppressing sensitivity to the spatial extent of the analysis window.

The renormalized Coleman bounce (solid black curves) provides an excellent description of the extracted bubble profiles and yields the smallest residuals for both $n_{\text{cut}}$ cases shown in Fig.~\ref{fig:profiles}. In particular, the corresponding weighted $\chi^2$ improves by about an order of magnitude relative to the bare classical profile and by over two orders of magnitude compared to the bare Coleman solution. While the renormalized classical bubble (solid gray curves) also captures the observed UV cutoff dependence, it exhibits systematic deviations in the outer tails of the profiles and is disfavored relative to the renormalized Coleman bounce by factors of a few in the weighted $\chi^2$. This preference cannot be attributed to our weighting scheme as it suppresses the most discrepant region of the renormalized classical profile, making the comparison conservative. We therefore conclude that the nucleation events observed in our real-time lattice simulations are best described by Euclidean instanton solutions in an appropriately renormalized effective potential. Results for intermediate values of $n_{\text{cut}}$ (not shown), discussed in detail in the companion paper~\cite{PRD_companion}, support the same qualitative conclusions.


\emph{Decay rates---}%
Using the best-fit $V_{\text{eff}}$ inferred from the bubble profile at each $n_{\text{cut}}$ (for the renormalized Coleman profile), we next solve Eq.~\eqref{eq:EOM} to obtain the corresponding bounce solution $\phi_b(x)$ and evaluate the exponent
\begin{equation}\label{eq:bounce_action}
B=S_E[\phi_b(x)]-S_E[\phi_{\text{FV}}],
\end{equation}
which represents the difference between the Euclidean action $S_E$ of the bubble configuration and that of the homogeneous false vacuum. This allows us to compute the corresponding decay rate predicted by the instanton formalism,
\begin{equation}\label{eq:decay_rate}
\frac{\Gamma}{L} =\bar{D} \times \mu^{2}\left(\frac{B}{2\pi}\right)e^{-B},
\end{equation}
where $\bar{D}$ is the dimensionless fluctuation determinant and $\mu$ sets the overall mass scale. Since $\bar{D}$ encodes one-loop corrections about the leading semiclassical contribution, it is expected to have the form $\bar{D}\sim 1+O(\hbar)$~\cite{braden_mass_2023}. Future work will aim to evaluate this determinant prefactor using methods similar to those implemented in Ref.~\cite{ekstedt_bubbledet_2023}, and compare the results with fluctuations measured around the stacked bubble profiles. Here, we follow Ref.~\cite{braden_mass_2023} and approximate $\bar{D} \simeq 1$, taking $\mu=m_0\sqrt{\lambda^2-1}$ to be the bare false-vacuum mass.

The resulting predictions for the logarithm of the decay rate $\Gamma$ are listed in the first column of Table~\ref{tab:decay_rates}, with uncertainties obtained by propagating the $1\sigma$ error on the best-fit value of $\Delta \tilde{V}$. We find that a higher UV cutoff leads to a larger decay rate, consistent with the behavior reported in Ref.~\cite{braden_mass_2023}. Qualitatively, this trend arises because mass renormalization decreases the curvature of the potential at the minima, thereby lowering the barrier and making bubble nucleation increasingly efficient as additional short-wavelength fluctuations are included. By contrast, the rate predicted from the bare Coleman bounce is extremely suppressed ($\ln (\Gamma) = -36.82$) and would not yield any observable nucleation events within typical simulation timescales.

\begin{table}[b]
\caption{\label{tab:decay_rates}%
Comparison between the decay rates extracted from simulations (third column)
and predicted using the Coleman bounce formalism with the best-fit effective potential (second column). The Euclidean prediction using the bare potential is $\ln(\Gamma)=-36.82$.}
\renewcommand{\arraystretch}{1.6}
\begin{ruledtabular}
\begin{tabular}{ccc}
$n_{\text{cut}}$ & Renormalized $\ln(\Gamma)$ & Measured $\ln(\Gamma)$ \\
\colrule
128  & $-9.66\pm0.06$ & $-7.50 \pm 0.04$ \\
256  & $-7.76 \pm 0.07$ & $-6.49 \pm 0.03$ \\
512 & $-5.89 \pm 0.02$ & $-5.62 \pm 0.03$ \\
1024 & $-4.15 \pm 0.03$ & $-4.55 \pm 0.03$ \\
\end{tabular}
\end{ruledtabular}
\end{table}

We then compare these instanton predictions with decay rates extracted directly from our real-time lattice simulations. For each UV cutoff, we evolve $1000$ independent realizations and compute the survival probability $P_{\mathrm{surv}}(t)$, defined as the fraction of realizations that have not yet nucleated by time $t$. The decay rate is obtained by fitting the logarithm of the survival probability with the affine form $\ln P_{\mathrm{surv}}(t)=-\Gamma (t-t_c)$, where both $\Gamma$ and the offset $t_c$ are treated as free parameters. Fits are performed until either 90\% of realizations have nucleated or five lattice crossing times have elapsed, whichever occurs first.
The resulting values of $\ln(\Gamma)$ are presented in the second column of Table~\ref{tab:decay_rates}, together with the corresponding $1\sigma$ fit uncertainties. 

While the extracted decay rates are exponentially larger than the bare prediction (as in previous studies~\cite{braden_erratum_2019,jenkins_analog_2024}), they are within the same order of magnitude as the Euclidean rates obtained from the renormalized effective potential, and exhibit the same increasing trend with the UV cutoff. The residuals $\Delta \ln (\Gamma) = \ln (\Gamma_{\text{sim}})-\ln (\Gamma_{\text{th}})$ decrease with increasing $n_{\text{cut}}$, changing from positive to slightly negative values, and remain of order $O(1)$ across all cutoffs. Factors of a few in the decay rate are consistent with the expected scale of the omitted one-loop fluctuation determinant~\cite{braden_mass_2023}; additional $O(1)$ corrections may also arise due to the relaxation of the initial statistical state of the field, which was shown to yield time-dependent nucleation rates in finite-temperature simulations~\cite{pirvu_thermal_2024,pirvu_thermal_2024_b}. A detailed investigation of these contributions, requiring an explicit computation of $\bar{D}$, is left for future work. Overall, numerical factors between simulation results and renormalized predictions for $\Gamma$ are small relative to the 12–14 orders of magnitude separating them from the bare theory, and are comparable in size to expected subleading corrections. Our findings therefore provide compelling evidence that renormalization resolves the apparent discrepancy between decay rates calculated in the Coleman instanton framework and those extracted from real-time lattice simulations.

\emph{Conclusions---}%
In this work, we have shown that the nucleation events observed in stochastic lattice simulations of vacuum decay can be brought into quantitative agreement with Euclidean instanton predictions by using an appropriately renormalized effective potential. In particular, ensemble-averaged bubble profiles are remarkably well-fitted by Coleman bounce solutions derived from such a potential; long-standing tensions between predicted and observed decay rates are then also resolved. Should this agreement persist beyond the specific class of potentials and dimensionality considered here, it would suggest that real-time simulations and Euclidean instantons are equivalent semiclassical descriptions of FVD. This would imply that interpreting FVD as the tunneling of a collective coordinate is equivalent to the classical evolution of vacuum fluctuations, whose statistics are encoded in the Coleman instanton formalism~\cite{braden_new_2019}.

To further test the robustness of these conclusions, future work will explore whether the observed correspondence generalizes to broader classes of scalar potentials and higher spatial dimensions. Since the renormalization effects described here arise through local modifications of the effective potential, we expect a similar qualitative mechanism to persist beyond 1+1d, although this remains to be investigated quantitatively. Establishing the extent of this validity regime will be important for connecting lattice studies of vacuum decay to cosmological settings.

The renormalization framework explored here will also facilitate direct comparisons between lattice simulations and forthcoming cold-atom analog experiments~\cite{fialko_fate_2015,braden_towards_2018,jenkins_generalized_2023}, which will provide a unique opportunity to probe the full nonperturbative dynamics of FVD. This will allow us to assess how accurately semiclassical methods capture the decay process and will guide the refinement of existing theoretical tools. The lattice approach may then provide new insight into real-time aspects of bubble nucleation, including possible oscillon precursors and clustering effects~\cite{pirvu_bubble_2022,pirvu_bubble_2023}, with potential implications for cosmological observables such as stochastic gravitational-wave backgrounds~\cite{kosowsky_gravitational_1992, kamionkowski_gravitational_1994,Caprini:2015zlo}.


\section{Acknowledgements}

\begin{acknowledgments}
We thank Dalila Pîrvu, Sergey Sibiryakov, Ian Moss, and Niayesh Afshordi for fruitful discussions. This work was supported by the Science and Technology Facilities Council through the UKRI Quantum Technologies for Fundamental Physics Programme (Grant No. ST/T005904/1). E.H. was supported by a Gates Cambridge Scholarship (grant OPP1144 from the Bill \& Melinda Gates Foundation). A.C.J. was supported by the Engineering and Physical Sciences Research Council through a Stephen Hawking Fellowship (Grant No. EP/U536684/1) and by a Gavin Boyle Fellowship at the Kavli Institute for Cosmology, Cambridge. M.C.J. is supported by a Discovery Grant from the Natural Sciences and Engineering Research Council of Canada. This research was supported in part by Perimeter Institute for Theoretical Physics. Research at Perimeter Institute is supported by the Government of Canada through the Department of Innovation, Science and Economic Development Canada and by the Province of Ontario through the Ministry of Research, Innovation and Science. Our simulations were performed on the Hypatia cluster at UCL, using computing equipment funded by the Research Capital Investment Fund provided by UKRI and partially funded by the UCL Cosmoparticle Initiative. We are grateful to Edd Edmondson for technical support. We acknowledge the use of the Python packages \texttt{NumPy}~\cite{harris_array_2020}, \texttt{SciPy}~\cite{virtanen_scipy_2020} and \texttt{matplotlib}~\cite{hunter_matplotlib_2007}.

Based on the CRediT (Contribution Roles Taxonomy) system, the authors contributed as follows. E.H. : conceptualization; methodology; software; formal analysis; investigation;
data curation; validation; visualization; and writing (original draft). A.C.J.: conceptualization; methodology; software; validation; and writing (review and editing). H.V.P.: conceptualization; methodology; validation; writing (review and editing); and project administration. A.P.: conceptualization; validation; and writing (review). M.C.J.: conceptualization; and writing (review). 
\end{acknowledgments}



\bibliography{bibliography}

\pagebreak
\section*{End Matter}

\subsection{Construction of the renormalized effective potential}

\begin{figure}[b]
    \centering
    \includegraphics[width=\columnwidth]{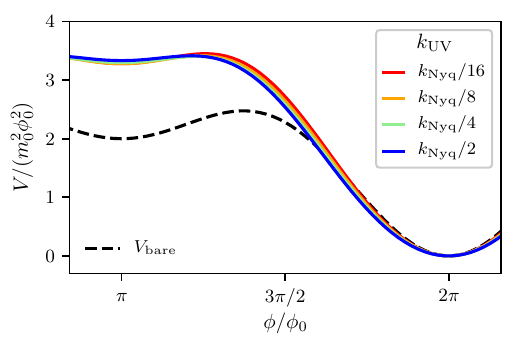}
    \caption{Effective potentials $V_\text{eff}$ [Eq.~\eqref{eq:V_eff}; colored lines] obtained by fitting the observed bubble profiles for 4 different values of the UV cutoff $k_\text{UV}$ against a renormalized Coleman template, with $k_\text{Nyq}$ representing the Nyquist wavevector. These are compared to the bare potential $V_\text{bare}$ [Eq.~\eqref{eq:Vbare}; dashed black line].}
    \label{fig:potential}
\end{figure}

Here we describe the Gaussian resummation procedure used to construct the renormalized effective potential $V_{\mathrm{eff}}$ from the bare lattice potential. Expanding the effective force $F_{\mathrm{eff}}(\bar{\phi})= -\langle V_{\text{bare}}'(\bar{\phi}+\delta\phi)\rangle \big|_{\bar{\phi}}$ into powers of the Gaussian fluctuations $\delta\phi$ and taking the ensemble average, we obtain an expression that depends only on the real-space variance
\begin{equation}
    \langle \delta\phi^2\rangle=\sigma_{\phi}^2=\frac{1}{2L}\sum_{n\neq 0}^{n_{\text{cut}}}\frac{1}{\sqrt{k_n^2+m_{\text{eff},i}^2}}.
\end{equation}
The effective true and false vacuum masses, defined such that $m^2_{\text{eff},i}=-\frac{dF_{\mathrm{eff}}}{d\bar{\phi}}\big|_{\bar{\phi_i}}$ with $i\in\{\text{TV, FV}\}$, are then determined by the nonlinear equation
\begin{equation}\label{eq:meff}
m^2_{\text{eff},i}
=m_0^2\left(s_i e^{-\frac{\sigma_{\phi}^2\left(m_{\text{eff},i}\right)}{2\phi_0^2}}+\lambda^2e^{-\frac{2\sigma_{\phi}^2\left(m_{\text{eff},i}\right)}{\phi_0^2}}\right),
\end{equation}
with $s_{\text{TV}}=1$ and $s_{\text{FV}}=-1$~\cite{braden_mass_2023}.

Although Eq.~\eqref{eq:meff} fixes the curvature of $V_{\text{eff}}$ near its local minima, the full shape of the renormalized potential between the true and false vacua is more challenging to calculate from first principles. Following Ref.~\cite{braden_mass_2023}, we model $V_{\text{eff}}$ by expanding it into a three-term cosine series 
\begin{equation}\label{eq:V_eff}
    V_{\text{eff}}(\phi)=\sum_{n=1}^3 c_n\left(m_{\text{eff,i}}, \Delta\tilde{V}\right) \left[1-\cos \left(n\frac{\phi}{\phi_0}\right)\right],
\end{equation}
where the coefficients $c_n$ are determined by the two effective masses and the dimensionless energy splitting $\Delta\tilde{V}$, treated here as a tunable parameter. Explicitly, we impose $V''_{\text{eff}}(\phi_{\text{FV}})=m_{\text{eff,FV}}^2$, $V''_{\text{eff}}(\phi_{\text{TV}})=m_{\text{eff,TV}}^2$, and $V_{\text{eff}}(\phi_{\text{FV}})-V_{\text{eff}}(\phi_{\text{TV}})=m_0^2\phi_0^2\Delta\tilde{V}$. This form preserves the (symmetry-protected) vacuum structure and periodicity of the bare potential, while providing a minimal parametrization of the residual freedom in the shape of $V_{\text{eff}}$ that is not fixed by local curvature constraints. We have verified that extending the expansion by an additional cosine mode does not significantly affect the bubble profiles or decay rates evaluated at the best-fit coefficient values.

Fig.~\ref{fig:potential} shows the bare potential $V_{\text{bare}}$, as well as the effective potentials obtained for each UV cutoff by fitting $\Delta\tilde{V}$ to the observed bubble profiles against a renormalized Coleman template.
It is interesting to note that these best-fit $\Delta\tilde{V}$ values, while significantly different from the bare value $\Delta\tilde{V}_\mathrm{bare}=2$, are insensitive to the UV cutoff (to within fit uncertainties) across the range of cutoffs considered here. This suggests that the renormalization flow in this $k_\mathrm{UV}$ regime is dominated by the renormalization of the true and false vacuum masses.

\subsection{Bubble profile extraction}

Our procedure for extracting bubble profiles from the simulations is as follows (further methodological details are provided in our companion paper~\cite{PRD_companion}). Bubbles are first identified in individual lattice realizations by locating connected near-true-vacuum regions that persist at all subsequent times, using a threshold well outside the typical amplitude of false-vacuum fluctuations. A finite spacetime window centered on the detected nucleation event is then selected for analysis, and realizations containing disconnected true-vacuum domains (secondary bubbles) within this area are discarded. Since bubbles generally nucleate with a nonzero center-of-mass velocity in the simulation frame, each configuration must be Lorentz boosted back into its approximate rest frame~\cite{pirvu_bubble_2022,pirvu_bubble_2023}. This is done by optimizing the boost velocity so that the full field evolution becomes maximally symmetric about the axis defined by the expanding true-vacuum region at late times. Deboosted bubbles are aligned in space by matching their symmetry axes and in time by synchronizing their late-time growth; they are then stacked to obtain an ensemble-averaged field configuration as shown in Fig.~\ref{fig:average_bubble}. For each value of $n_{\text{cut}}$, we include 500 independent realizations in the stack.

The nucleation time of the average bubble is determined by tracking outer wall trajectories and identifying the onset of hyperbolic expansion, after which  $x_{\text{wall}}(t\geq t_{\text{nuc}})=x_{\text{sym}}\pm\sqrt{R^2+(t-t_{\text{nuc}})^2}$. We infer $t_{\text{nuc}}$ and $R$ for 10 field-amplitude thresholds as the bubble wall position is not uniquely defined beyond the idealized thin-wall limit. This threshold range was chosen empirically such that the corresponding isocontours remain approximately at rest prior to $t_{\text{nuc}}$ (shaded regions in Fig.~\ref{fig:average_bubble}). The final nucleation time is taken as the mean of these threshold estimates, and the bubble profile is extracted from the corresponding slice of the ensemble-averaged field. Statistical uncertainties are estimated through bootstrap resampling of the stack.

\end{document}